\documentclass[aps,twocolumn,pra,superscriptaddress,amsmath,showpacs,tightenlines]{revtex4}
\usepackage[dvips]{graphicx}
\usepackage{epsfig}
\usepackage{color}
\usepackage[normalem]{ulem}

\begin{document}
\title{Implementation of standard quantum error correction codes for solid-state qubits}

\author{Tetsufumi Tanamoto}
\affiliation{Corporate R \& D center, Toshiba Corporation,
Saiwai-ku, Kawasaki 212-8582, Japan}

\date{\today}

\begin{abstract}
In quantum error-correcting code (QECC), many quantum operations 
and measurements are necessary to correct errors in logical qubits. 
In the stabilizer formalism, which is widely used in QECC, generators $G_i (i=1,2,..)$ consist of multiples of Pauli 
matrices and perform encoding, decoding and measurement. 
In order to maintain encoding states, the stabilizer Hamiltonian $H_{\rm stab}=-\sum_i G_i$
is suitable because its ground state corresponds to the code space.
On the other hand, Hamiltonians of most solid-state qubits have two-body interactions 
and show their own dynamics. In addition solid-state qubits are fixed on substrate
and qubit-qubit operation is restricted in their neighborhood.
The main purpose of this paper is to show how to directly generate the stabilizer Hamiltonian $H_{\rm stab}$ from 
conventional two-body Hamiltonians with Ising interaction and $XY$ interaction
by applying a pulse control method such as an NMR technique.
We show that generation times of $H_{\rm stab}$ for nine-qubit code, 
five-qubit code and Steane code are estimated to be less than 300~ns when typical experimental data of 
superconducting qubits are used, and sufficient pulse control is assumed.
We also show how to prepare encoded states from an initial state $|0....0\rangle$.
In addition, we discuss an appropriate arrangement of 
two- or three-dimensional arrayed qubits. 
\end{abstract}
%
\maketitle

\section{Introduction}
Similar to the digital computer,  
a rigid error-correcting system is required in the quantum computer. 
Various quantum error-correcting codes (QECC) have been developed
such as the standard code
~\cite{Shor,Steane,Gottesman,Nielsen,Steane2,Knill1,Knill2,Reed,Moussa},
the subsystem codes~\cite{Kribs,Poulin,Bacon}, 
and the topological code~\cite{Kitaev1,Bombin,Bravyi,Raussendorf,Milman,Fowler}.
In QECC, it is necessary for many qubits to be coherently entangled for constructing logical qubits.
For instance, nine qubits are required for a logical qubit of the nine-qubit code~\cite{Shor}, 
seven qubits are required for the seven-qubit Steane code, which is the smallest 
code of the general CSS code~\cite{Steane}, and so on~\cite{Gottesman,Nielsen}. 
In any quantum codes, many operations and measurements are required for encoding,
decoding and error-correcting processes.
There are strict requirements concerning the maximum error rate 
for the success of QECC~\cite{Gottesman,Nielsen,Steane2,Fowler}. 
All manipulations of many qubits 
should be done sufficiently within the coherence time.

In general, it is difficult to produce desired encoded states consisting of many qubits.
However, it is also difficult 
to maintain each entangled state during the time required in a flow of quantum computation~\cite{Knill1,Knill2,Reed,Moussa}. 
This problem arises when the encoded state is not the eigenstate of a system Hamiltonian.
The encoded state changes following the dynamics of the system Hamiltonian.
Assume that a computer system consists of many blocks. 
Each block must correlate with every other block
to carry out a definite set of quantum computations. 
As a simple structure of a computing system,
let us consider a system in which operations are synchronized to a system clock,  
which is the case with the present widely-used digital computers.
Then, all operations are carried out step by step 
as the system clock ticks the system time. 
Entangled states produced by CNOT gates or other quantum gates
appear only periodically when the entangled states are not the 
ground states of the system Hamiltonian. 
In such case, if each block of a system includes an individual entangled state,
it will be difficult to control the synchronization of the total system 
because the period of the desired entangled states 
differs depending on the dynamics of each block. 
Thus, it will be desirable for encoded states to be 
the ground states of Hamiltonians of the blocks.
Moreover, because each block of a system changes its role as system time passes, 
it is desirable that the Hamiltonian of each block 
changes depending on each calculation step.

In this paper, we show how to efficiently implement standard QECC in
solid-state qubit systems with natural two-body interactions, 
focusing on the stabilizer formalism. 
Stabilizer operators $\{G_i| 1\le i\le l\}$ are mutually commuting operators 
given by products of multiple Pauli matrices~\cite{Gottesman,Nielsen}. 
Logical qubit states are encoded 
into a mutual eigenspace $\mathcal{H}_S$ of dimension $2^l$ of these operators 
through measurements. For $l$ different stabilizers and $n$ physical qubits, a 
maximum number of $k = n - l$ logical qubits can be encoded into $\mathcal{H}_S$, 
whereas $k < n - l$ in case of subsystem encoding~\cite{Kribs,Poulin,Bacon}.
Although preparation of some ``quantum memory" blocks to where logical qubit 
states can be transferred or teleported is one solution to preserve logical qubit states, 
we consider that it is better to change a system Hamiltonian 
into a stabilizer Hamiltonian defined by $H_{\rm stab}\equiv -\sum_i G_i$,
because transformation or teleportation of encoded states requires more complexity.
We would also like to show how to generate encoded states without measurements.
The encoded states are generated by using operators that are modified from stabilizer operators.
Therefore, in this paper, we mainly describe the generation process of $H_{\rm stab}$.

In previous papers~\cite{stab,qecc-basel}, we showed that we can construct $G_i$ 
one by one based on the two-body Hamiltonian by using 
the appropriate pulse sequence. 
However, it is much more efficient to directly produce $H_{\rm stab}$. 
In this paper, we show how to directly create $H_{\rm stab}$ 
starting from the two-body Hamiltonian. 
$H_{\rm stab}$ has a complicated form of multiplied Pauli matrices.
We show that appropriate pulse sequences to generate 
$H_{\rm stab}$ can be found by inversely tracing a transformation from $H_{\rm stab}$ into single-qubit Hamiltonian.
We show that the direct creation of $H_{\rm stab}$ greatly reduces
the number of operations compared with our previous method in Ref.~\cite{stab,qecc-basel}.
This reduction is remarkable in the case of qubits with $XY$ interaction.
For example, the number of single-qubit rotations $N_{\rm rot}$ and that of 
qubit-qubit $XY$ interaction $N_{\rm int}$ are reduced from $N_{\rm rot}=44$ to $N_{\rm rot}=20$, 
and from $N_{\rm int}=288$ to $N_{\rm int}=132$,
respectively, for the Steane code. Similar results are 
obtained for the nine-qubit code and the five-qubit code.
Accordingly, operation time can also be reduced.
If we use a typical experimental parameter of superconducting qubits,
we can reduce the time required to generate $H_{\rm stab}$ by 48.4~\% ( 194~ns), 59.1~\% (127.5~ns) and 54.4~\% (257~ns) 
for the nine-qubit code, the five-qubit code and the Steane code, respectively.
The present method has the advantage that, as pulse control technology progresses,  
pulse error rate and speed are improved. 
Pulse errors can be corrected by using NMR techniques such as the composite-pulse 
method~\cite{Ernst,Haffner,Molmer,Hill,Torosov}, 
and the speed is increased by improving a control system operated by a digital computer.

\begin{figure}
\includegraphics[width=7cm,clip=true]{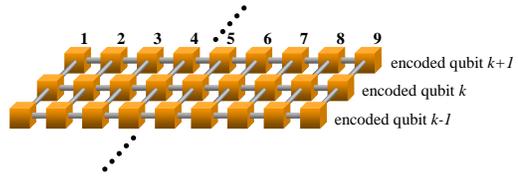}
\caption{Two-dimensional qubit array aiming at Shor's nine-qubit code. Boxes show qubits and bars 
between the boxes show interactions between qubits. Horizontal qubits constitute logical qubits.
} 
\label{qubitlattice}
\end{figure}

We also investigate a possible architecture of standard codes for solid-state qubits on lattice sites.
In general, interactions between solid-state qubits are 
restricted to their nearest-neighbor or next nearest-neighbor sites~\cite{Daniel,Yamamoto,You,Loss,Petta,Kane,tana}.
In order to prevent unexpected external noise, 
it is preferable for physical qubits in a logical 
qubit to be placed compactly in a small region. 
Moreover, for logical qubits to interact effectively with one another,
it is desirable to place logical qubits side by side.
Therefore, it is natural to construct logical qubit by one-dimensional (1D) qubit arrays 
and place them parallel as shown in Fig.~\ref{qubitlattice}.
In addition, frequent measurements in QECC require other qubit arrays for measurements. 
We will discuss possible setups of a qubit system.

As a general case, we consider always-on Hamiltonian in this paper. 
We think that we can separate logical qubits by effectively eliminating 
qubit-qubit interaction through the use of appropriate pulse sequences.

This paper is organized as follows:
In Sec.~\ref{sec:form} we establish the general procedure of generating the stabilizer Hamiltonian.
In Sec.~\ref{sec:code}, we show examples of generating 
the stabilizer Hamiltonian in the standard code, 
and in Sec.~\ref{sec:initial} we show how to
generate the code state. 
Finally, in Sec.~\ref{sec:architecture}, 
we consider possible qubit architecture realized by 
solid-state qubits. We close with a summary and conclusions in
Sec.~\ref{sec:conclusion}.

\section{Stabilizer Hamiltonian generation method}\label{sec:form}
\subsection{Stabilizer coding and stabilizer Hamiltonian}\label{sec:review}
In the stabilizer code~\cite{Gottesman,Nielsen}, 
encoding, decoding and error-correction are carried out based on the stabilizers,
which are mutually commutable and can be expressed by the Pauli matrices:
\begin{equation}
G_l=\otimes_{i=1}^{n} (X_i)^{x_i(G_l)}(Z_i)^{z_i(G_l)}
\label{ggg}
\end{equation}
($x_i(G_l), \ z_i(G_l) \in \{0,1\}$), where 
Pauli matrices are given by
\begin{equation}
X=
\left(
\begin{array}{cc}
0 & 1 \\
1 & 0 \\
\end{array}
\right), \
Y=
\left(
\begin{array}{cc}
0 & -i \\
i & 0 \\
\end{array}
\right), \ 
Z=
\left(
\begin{array}{cc}
1 & 0 \\
0 & -1 \\
\end{array}
\right).
\end{equation}
The codeword $|\Psi_m\rangle$ obeys the eigenvalue equation 
\begin{equation}
G_l|\Psi_m \rangle =|\Psi_m \rangle
\end{equation}

Conventionally, 
in order to construct encoding states, starting from an initial state $\Pi_{i=1}^k|0\rangle_i$, 
measurements over stabilizer operators of the selected code are repeated.
Depending on the measurement outcome, 
the common eigenstate is fixed to be the desirable encoded state.  
The correction procedure for the stabilizer code is 
carried out by measuring all relevant stabilizer operators. 

The stabilizer Hamiltonian $H_{\rm stab}$ is defined by
\begin{equation}
H_{\rm stab}=-\sum_{i=1}^{l}G_i, 
\label{totalstab}
\end{equation}
where the summation is taken over the constituent stabilizers of each code.
Owing to the commutability of the stabilizers $G_i$, 
the ground state of Eq.(\ref{totalstab}) is a common eigenstate of the 
stabilizers, which is the encoded logical state. 
For the sake of simplicity, we consider the standard codes without considering subsystem code
($k=n-l$).

\subsection{System Hamiltonian} 
The solid-state Hamiltonian controlled by pulse signals 
can be written as~\cite{You2,Regetti,tana2001}
\begin{equation}
H(t)=\sum_i \left[ \Omega_{0i} Z_i + 2\Omega_i \cos (\omega^{\rm rf}_i t+\delta_i) X_i\right] +\sum_{i<j}J_{ij}X_iX_j,
\label{app1}
\end{equation}
where $\Omega_i$ and $\omega^{\rm rf}_i$ are an amplitude and a frequency of a controlled signal 
applied to a qubit $i$.
If we move to a frame rotating with the radio-frequency $\omega^{\rm rf}_i$ about the z-axis,
$H^r =R^{-1} H(t)R$,
with $R=\exp [-i \sum_i (\omega^{\rm rf}_i t/2) Z_i]$, then the transferred static Hamiltonian 
$H'=H^r -\sum_i  (\omega^{\rm rf}_i t/2) Z_i$ is approximately given by
\begin{eqnarray}
H' &=&\sum_i \left[  \left(\Omega_{0i} - \frac{\omega^{\rm rf}_i}{2}\right) Z_i 
+\Omega_i (\cos \delta_i X_i -\sin \delta_i Y_i) \right]
\nonumber \\
&+&\sum_{i<j}\frac{J_{ij}}{2}
[X_iX_j +Y_iY_j].
\label{app2}
\end{eqnarray}
(high-frequency components $2\omega^{\rm rf}_i$ can be neglected). 
If Eq.~(\ref{app1}) includes an interaction of $\sum_{i<j}J_{ij}Z_iZ_j$
 instead of $\sum_{i<j}J_{ij}X_iX_j$, 
the final Hamiltonian Eq.~(\ref{app2}) includes the Ising interaction. 
The x-pulse and y-pulse for qubit $i$ are realized when 
$\delta=0$ and $\delta=-\pi/2$ signals are respectively 
applied to the qubit with $\omega^{\rm rf}_i=2\Omega_{0i}$.
We assume that each pulse is sufficiently strong for
interactions between qubits to be neglected during 
the pulse sequences ($\Omega_i,\Omega_{0i} > J_{ij}$).

Then the qubit Hamiltonian 
in the rotating frame of $\omega^{\rm rf}_i=2\Omega_{0i}$
is expressed by $H_q=H_0+H_{\rm int}$ 
where a single-qubit part $H_0$ is given by 
\begin{equation}
H_0= \sum_i H_{0i}= \sum_i \Omega_i X_{i}
\label{H0}
\end{equation}
The interacting part 
$H_{\rm int} =\sum_{ij}H_{\rm int}^{ij}$ is expressed by
\begin{equation}
H_{XY}=\sum_{i<j}H_{XY}^{ij}=\sum_{i<j}J_{ij}[X_i X_j+Y_iY_j],
\end{equation}
for $XY$ interaction, and
\begin{equation}
H_{\rm Ising}=\sum_{i<j}H_{\rm Ising}^{ij}=\sum_{i<j}J_{ij}Z_i Z_j,
\label{Hising}
\end{equation}
for Ising interaction.

\subsection{Dynamic generation of stabilizer Hamiltonian} 
The generation of $H_{\rm stab}$ from  $H_q$ consists of two steps.
The first step is to extract a
single-qubit part or a pure two-body interaction part from a qubit Hamiltonian $H_q$.
The second step is to construct $H_{\rm stab}$ dynamically with pulse sequences 
by using a selected single qubit part $H_{\rm ini}$ and qubit-qubit interactions $H_{\rm int}^{ij}$.
Because the second step is the core framework of this paper, 
we first describe the second step of dynamical transformation 
to  $H_{\rm stab}$. The extraction method is described in the next section~\ref{sec:extract}.
 
The transformation from the two-body Hamiltonian $H_q$ to the many-body 
Hamiltonian $H_{\rm stab}$ is carried out 
dynamically by using a time evolution of a system  
starting from a simple initial Hamiltonian $H_{\rm ini}\propto X_i$, $Y_i$, or
$Z_i$ ~\cite{stab,qecc-basel}. The time evolution of the generation process is illustrated
with the schematic notation $\rho(0) \stackrel{t H}{\longrightarrow } \rho(t)$,
where $\rho(t) = \exp(-iHt) \rho(0) \exp(iHt)$ is the density matrix for a
time-independent Hamiltonian $H$, or for an effective $H$ in the sense of the
average Hamiltonian theory~\cite{Ernst}. After the application of mutually inverse, unitary
operations according to 
\begin{equation}
\rho(0) \stackrel{\tau_{\rm op} H_{\rm op}
}{\longrightarrow } \ \ \stackrel{ \tau_{\rm ini} H_{\rm ini} }{\longrightarrow
} \ \ \stackrel{ -\tau_{\rm op} H_{\rm op} }{\longrightarrow } \rho(\tau_{\rm
ini}+2\tau_{\rm op}), 
\label{rho}
\end{equation}
the system evolves as if propagated by the effective
Hamiltonian $\exp(-i \tau_{\rm op} H_{\rm op}) H_{\rm ini} \exp(i \tau_{\rm op}
H_{\rm op})$ for a time $\tau_{\rm ini}$~\cite{stab}. 

To build $H_{\rm stab}$ from $H_{\rm ini}$, 
we need two elementary transformations: one
that rotates arbitrary single-qubit terms through an angle of $\pi/2$ and
another that increases the order of Pauli-matrix terms by one. 
Higher-order products of Pauli
matrices can be generated using the following transformations~\cite{stab}:
\begin{eqnarray}
\! e^{-i\theta [XY]_{12}} X_{1} e^{i\theta [XY]_{12}} \!\!\!&\!=\!&\!\! \cos (2\theta) X_{1} -\sin (2\theta) Z_{1} Y_{2}
\label{XYa}, \\
\! e^{-i\theta [XY]_{12}} Y_{1} e^{i\theta [XY]_{12}} \!\!\!&\!=\!&\!\! \cos (2\theta) Y_{1} +\sin (2\theta) Z_{1}X_{2} 
\label{XYb}, \\
\! e^{-i\theta [XY]_{12}} Z_{1} e^{i\theta [XY]_{12}} \!\!\!&\!=\!&\!\! \cos^2 (2\theta) Z_{1} +\sin^2 (2\theta) Z_{2}  \nonumber\\
\!\!\!&\!+\!&\!\!\frac{1}{2}\sin (4\theta)
[X_{1}Y_{2}-Y_{1}X_{2}]\:, 
\label{XYc} 
\end{eqnarray}
for $XY$ interaction. 
When $\theta=\pi/4$ we can change the number of Pauli matrices given by
\begin{eqnarray}
X_{1} &\rightarrow &  -Z_{1} Y_{2}, \label{XYa1} \\ 
Y_{1} &\rightarrow &   Z_{1} X_{2}, \label{XYb1} \\
Z_{1} &\rightarrow &   Z_{2}. \label{XYc1} 
\end{eqnarray}
For Ising interaction, we use the relations given by
\begin{eqnarray}
& & e^{-i\theta Z_{1}Z_{2} } X_{1} e^{i\theta Z_{1}Z_{2}} 
\!=\! \cos (2\theta) X_{1} +\sin (2\theta) Y_{1} Z_{2},
\label{ZZa}\\
& & e^{-i\theta Z_{1}Z_{2} } Y_{1} e^{i\theta Z_{1}Z_{2}} 
\!=\! \cos (2\theta) Y_{1} -\sin (2\theta) X_{1} Z_{2}\:,
\label{ZZb}
\end{eqnarray}
Then, for $\theta=\pi/4$, we can 
change the number of Pauli matrices given by
\begin{eqnarray}
& & X_{1} \rightarrow   Y_{1} Z_{2}, \label{ZZa1} \\
& & Y_{1} \rightarrow  -X_{1} Z_{2}, \label{ZZb1}\\
& & Z_{1} \rightarrow  Z_{1}. \label{ZZc1}
\end{eqnarray}
By combining these equations with single-qubit rotations, we can change $H_q$ to 
$H_{\rm stab}$.

\subsection{Extracting $H_{\rm ini}$ and $H_{\rm op}$ from a qubit Hamiltonian}\label{sec:extract}
In order to use the above-mentioned dynamic method, the important step is 
to extract a
single-qubit part or a pure two-body interaction part from a qubit Hamiltonian $H_q$. 
This process is carried out using the Baker-Campbell-Hausdorff (BCH) formula
\cite{Ernst}. 
Here, we assume that qubits interact with their nearest-neighbor qubits. 
Then, in order to define a logical qubit, we have to determine the locations of physical qubits in a logical qubit.
In this section, after we explain the BCH formula, we would like to 
define a logical qubit arranged on lattice sites. Then, finally we will show 
how to extract a single-qubit part $H_{\rm ini}$ 
and a pure two-body interaction $H_{\rm op}$ from 
the Hamiltonian of a qubit lattice.

\subsubsection{Manipulation by using the Baker-Campbell-Hausdorff (BCH) formula}
A desirable part of the original Hamiltonian $H_q$ 
is extracted by using appropriate pulse sequences~\cite{stab}.
The basic idea can be illustrated by using the standard NMR Hamiltonian $H_{\rm nmr}=\sum_i
\varepsilon_i Z_i+ \sum_{i<j} J Z_i Z_j$. 
In this case, because of the property  $[H_0, H_{\rm int}]=0$, $H_0$ and
$H_{\rm int}$ can be separately obtained by using a simple pulse sequence. 
The interaction part $H_{\rm Ising}$ can be extracted by using two 
sandwiched $\pi$-pulses such as 
$\exp(i\tau H_{\rm Ising})=e^{-i(\pi/2) \sum_j Y_j} e^{ i(\tau/2) 
H_{\rm nmr}} e^{ i(\pi/2) \sum_j Y_j} e^{ i(\tau/2) H_{\rm nmr}}$. 
For the general Hamiltonian 
(Eqs.(\ref{H0})-(\ref{Hising})), because $[H_0, H_{\rm int}]\neq 0$, 
we approximately obtain a desirable part by repeatedly applying the Baker-Campbell-Hausdorff (BCH) formula.
For $A=h_a+h_b$ (original
Hamiltonian) and $B=h_a-h_b$ (transferred by applying a $\pi$ pulse)
with $h_a=i\tau H_a$ and $h_b=i\tau H_b$, 
we can extract $h_a=i\tau H_a$ by using the relation given by
\begin{equation}
(e^Ae^B)^n \approx \exp( i2 t_0 H_a + (t_0^2/n)[H_a,H_b])
\label{AB}
\end{equation} 
($t_0\equiv n\tau$). Thus, as long as $(t_0/n)||H_b|| \ll 1$ where
$||A||=[\mathrm{Tr}(A^\dagger A)/d]^{1/2}$ is the standard operator norm in a
Hilbert space of dimension $d$, we can neglect the second term. 
As the number $n$ of repetitions increases, this approximation improves. 

In the following sections, we use an extended form of Eq.(\ref{AB}) described by
\begin{eqnarray}
\lefteqn{ (e^Ae^B  e^{B'}e^{A'})^n \approx [\exp( 2h_a + [h_b,h_a]) \exp( 2h_a' - [h_b',h_a'])]^n } \nonumber \\
      & \approx& \exp( 2n(h_a+h_a') + n[h_b,h_a]- n[h_b',h_a'] +4 n[h_a, h_a'] ), 
\nonumber \\
\label{ABAB}
\end{eqnarray}
where $A'=h_a'+h_b'$ and $B'=h_a'-h_b'$. $2(h_a+h_a')$ is the target Hamiltonian.
In the following two subsections, we show how to extract a desirable interaction 
term $H_{\rm int}^{ij}$ and a single-qubit part $H_0$ from $H_q$ by using 
Eq.~(\ref{ABAB}).

\subsubsection{Qubit lattice and logical qubit}\label{sec:lattice}
We consider a qubit lattice in which physical qubits are arrayed on a lattice site 
interacting with their neighboring qubits.
The simplest arrangement is a 1D array as shown in Fig.~\ref{qubitlattice}. 
Then we can interact logical qubits with their nearest-neighbor logical 
qubits by using interactions between physical qubits. 
The number of qubits in each 1D array depends on how many physical qubits 
are required to construct a single logical qubit.
In Fig.~\ref{qubitlattice}, nine qubits constitute a logical qubit.

\subsubsection{Selection of a single-qubit Hamiltonian}\label{sec:extractH_0}
Here we show how to extract $H_0$ from $H_q$ for 2D qubit lattice, 
assuming always-on interactions between qubits. 
As an example, we consider logical qubits consisting of five qubits.
In a 1D qubit array, $H_0$ is obtained by choosing 
$B=-h_1+h_2-h_3+h_4-h_5-\sum_i h_{ij}$ and
$B'=h_1-h_2+h_3-h_4+h_5-\sum_i h_{ij}$ while 
$A$ and $A'$ are $H_q$ in Eq.~(\ref{ABAB})
[$h_i=\tau H_{0i}$ and $h_{ij}=\tau H_{\rm int}^{ij}$]. 
This procedure can be extended to the 2D lattice case 
by taking into account interactions between different logical qubits.

In this section, we treat Hamiltonians that include two types of Pauli matrices or fewer 
such as Eq.~(\ref{app1}) or Eq.~(\ref{app2}) with $\omega_i^{\rm rf}=2\Omega_{0i}$.
For Eq.~(\ref{app1}), `$\pi$-pulse' corresponds to $\pi$-pulse around $y$-axis.
For Eq.~(\ref{app2}) with $\omega_i^{\rm rf}=2\Omega_{0i}$, `$\pi$-pulse'
corresponds to $\pi$-pulse around $z$-axis, which can also be produced by 
$\pi$-pulse around $y$-axis after that around $x$-axis.
Extraction of $H_0$ and two-body interaction from the Hamiltonian Eq.~(\ref{app2}) 
with $\omega_i^{\rm rf}\neq 2\Omega_{0i}$ 
is described in Appendix~\ref{appendixA}.

The 2D lattice Hamiltonian is given by 
\begin{equation}
H^{2D}=\sum_k H_q^{(k)},
\end{equation}
where
\begin{equation}
H_q^{(k)}=H_0^{(k)}+H_{\rm int}^{(k)}+H_{\rm int}^{(k,k+1)}.  
\end{equation}
$H_{\rm int}^{(k,k+1)}$ shows an interaction term between $k$-th logical qubits and $k+1$-th qubits.
In order to separate different logical qubits, $H_{\rm int}^{(k,k+1)}$ should be erased.
We apply 
$\pi$-pulses to 
(i) qubits 1,3,5 of ...,$k-1$-th, $k+1$-th, ... arrays for $A$,
(ii) qubits 1,3,5 of ...,$k$-th, $k+2$-th, ... arrays for $B$,
(iii) qubits 2,4  of ...,$k-1$-th, $k+1$-th, ...  arrays for $B'$,
and (iv) qubits 2,4 of qubits of ...,$k$-th, $k+2$-th, ...  arrays for $A'$:
\begin{eqnarray}
\lefteqn{A = \cdots} \nonumber \\ 
  &-&h_1^{(k-1)}+h_2^{(k-1)}-h_3^{(k-1)}+h_4^{(k-1)}-h_5^{(k-1)} -h_{\rm int}^{(k-1)}
 \nonumber \\
  &-& h_{11}^{(k-1,k)}+h_{22}^{(k-1,k)}-h_{33}^{(k-1,k)}+h_{44}^{(k-1,k)}
     -h_{55}^{(k-1,k)}  \nonumber \\ 
%
 &+& h_q^{(k)}
 \nonumber \\
 &-&  h_{11}^{(k,k+1)}+h_{22}^{(k,k+1)}-h_{33}^{(k,k+1)}+h_{44}^{(k,k+1)}
     -h_{55}^{(k,k+1)}  \nonumber \\ 
  &-&h_1^{(k+1)}+h_2^{(k+1)}-h_3^{(k+1)}+h_4^{(k+1)}-h_5^{(k+1)} -h_{\rm int}^{(k+1)}
 \nonumber \\
%
& & \cdots 
\end{eqnarray}
\begin{eqnarray}
\lefteqn{B = \cdots +h_q^{(k-1)} } \nonumber \\ 
 &-& h_{11}^{(k-1,k)}+h_{22}^{(k-1,k)}-h_{33}^{(k-1,k)}+h_{44}^{(k-1,k)}
     -h_{55}^{(k-1,k)}  \nonumber \\ 
 &-&h_1^{(k)}+h_2^{(k)}-h_3^{(k)}+h_4^{(k)}-h_5^{(k)} -h_{\rm int}^{(k)}
 \nonumber \\
 &-&  h_{11}^{(k,k+1)}+h_{22}^{(k,k+1)}-h_{33}^{(k,k+1)}+h_{44}^{(k,k+1)}
     -h_{55}^{(k,k+1)}  \nonumber \\ 
 &+& h_q^{(k+1)}  \cdots
\end{eqnarray}
\begin{eqnarray}
\lefteqn{B' = \cdots} \nonumber \\ 
  &+&h_1^{(k-1)}-h_2^{(k-1)}+h_3^{(k-1)}-h_4^{(k-1)}+h_5^{(k-1)} -h_{\rm int}^{(k-1)}
 \nonumber \\
  &+& h_{11}^{(k-1,k)}-h_{22}^{(k-1,k)}+h_{33}^{(k-1,k)}-h_{44}^{(k-1,k)}
     +h_{55}^{(k-1,k)}  \nonumber \\ 
 &+& h_q^{(k)}
 \nonumber \\
 &+&  h_{11}^{(k,k+1)}-h_{22}^{(k,k+1)}+h_{33}^{(k,k+1)}-h_{44}^{(k,k+1)}
     +h_{55}^{(k,k+1)}  \nonumber \\ 
  &+&h_1^{(k+1)}-h_2^{(k+1)}+h_3^{(k+1)}-h_4^{(k+1)}+h_5^{(k+1)} - h_{\rm int}^{(k+1)}
 \nonumber \\
%
& & \cdots 
\end{eqnarray}
\begin{eqnarray}
\lefteqn{A' = \cdots  +h_q^{(k-1)} } \nonumber \\ 
 &+& h_{11}^{(k-1,k)}-h_{22}^{(k-1,k)}+h_{33}^{(k-1,k)}-h_{44}^{(k-1,k)}
     +h_{55}^{(k-1,k)}  \nonumber \\ 
 &+&h_1^{(k)}-h_2^{(k)}+h_3^{(k)}-h_4^{(k)}+h_5^{(k)} -h_{\rm int}^{(k)}
 \nonumber \\
 &+&  h_{11}^{(k,k+1)}-h_{22}^{(k,k+1)}+h_{33}^{(k,k+1)}-h_{44}^{(k,k+1)}
     +h_{55}^{(k,k+1)}  \nonumber \\ 
 &+& h_q^{(k-1)} \cdots 
\end{eqnarray}
where  $h_q^{(k)}=\tau (H_0^{(k)}+H_{\rm int}^{(k)})$. 
By using Eq.(\ref{ABAB}), we obtain $H_{\rm eff}=2 \sum_k H_0^{(k)}$.

\subsubsection{Selection of two-body interaction}\label{sec:extractH_int}
Next, we show how to extract the interaction term 
$H_{\rm int}^{ij}$ between two qubits  
in order to use Eqs.(\ref{XYa})-(\ref{XYc}) or Eqs.(\ref{ZZa})-(\ref{ZZb})
for the 2D lattice qubits.
As an example, we consider a case of extracting $h_{23}=i\tau H_{\rm int}^{23}$ in 
five-qubit array. 
The required transformation is given by extending the results of Ref.~\cite{qecc-basel}.
$A$ in Eq.(\ref{ABAB}) is the original Hamiltonian such as $A=\tau (H_0 + H_{\rm int})$. 
$B$ in Eq.(\ref{ABAB}) is given by applying $\pi$ pulse to qubits 2,3,5 of $(k+2n)$-th logical qubits 
and qubits 1,4 of $(k+2n-1)$-th logical qubits ($n$ is an integer):
\begin{eqnarray}
\lefteqn{B= \cdots} \nonumber \\ 
 &-& h_1^{(k-1)}+h_2^{(k-1)}+h_3^{(k-1)}-h_4^{(k-1)}+h_5^{(k-1)} 
 \nonumber \\
 & & -h_{12}^{(k-1)}+h_{23}^{(k-1)}-h_{34}^{(k-1)}-h_{45}^{(k-1)}
     -h_{\rm int}^{(k-1,k)}  \nonumber \\ 
 &+&h_1^{(k)}-h_2^{(k)}-h_3^{(k)}+h_4^{(k)}-h_5^{(k)} 
 \nonumber \\
 & & -h_{12}^{(k)}+h_{23}^{(k)}-h_{34}^{(k)}-h_{45}^{(k)}
     -h_{\rm int}^{(k,k+1)}  \nonumber \\ 
 &-&h_1^{(k+1)}+h_2^{(k+1)}+h_3^{(k+1)}-h_4^{(k+1)}+h_5^{(k+1)} 
 \nonumber \\
 & & -h_{12}^{(k+1)}+h_{23}^{(k+1)}-h_{34}^{(k+1)}-h_{45}^{(k+1)}
     -h_{\rm int}^{(k+1,k+2)}  \nonumber \\ 
& & \cdots 
\end{eqnarray}
where $h_i^{(k)} \equiv i\tau H_0^{(k)}$ and $h_{ij}^{(k)}=i\tau H_{\rm int}^{(k)}$. 
$B'$ is given by applying $\pi$ pulse to qubits 2,3,5 of $(k+2n-1)$-th logical qubits 
and qubits 1,4 of $(k+2n)$-th logical qubits ($n$ is an integer):
\begin{eqnarray}
\lefteqn{B'= \cdots} \nonumber \\ 
 &+&h_1^{(k-1)}-h_2^{(k-1)}-h_3^{(k-1)}+h_4^{(k-1)}-h_5^{(k-1)} 
 \nonumber \\
 & & -h_{12}^{(k-1)}+h_{23}^{(k-1)}-h_{34}^{(k-1)}-h_{45}^{(k-1)}
     -h_{\rm int}^{(k-1,k)}  \nonumber \\ 
 &-& h_1^{(k)}+h_2^{(k)}+h_3^{(k)}-h_4^{(k)}+h_5^{(k)} 
 \nonumber \\
 & & -h_{12}^{(k)}+h_{23}^{(k)}-h_{34}^{(k)}-h_{45}^{(k)}
     -h_{\rm int}^{(k,k+1)}  \nonumber \\ 
 &+&h_1^{(k+1)}-h_2^{(k+1)}-h_3^{(k+1)}+h_4^{(k+1)}-h_5^{(k+1)} 
 \nonumber \\
 & & -h_{12}^{(k+1)}+h_{23}^{(k+1)}-h_{34}^{(k+1)}-h_{45}^{(k+1)}
     -h_{\rm int}^{(k+1,k+2)}  \nonumber \\ 
& & \cdots 
\end{eqnarray}
The $A'$ is obtained by applying $\pi$ pulse to all qubits given by
\begin{equation}
A'=\tau (-H_0 +H_{\rm int}).
\end{equation}
By using Eq.(\ref{ABAB}), we can obtain $\sum_k 4h_{23}^{(k)}$.

The perturbation terms in Eq.(\ref{ABAB})
are described in Appendix~\ref{appendixB}.
For the selection of $\sum_k 4h_{23}^{(k)}$,
the perturbation is estimated as  
$|| H_{\rm pert} || \approx 10 \tau N_{\rm qubit} J\Omega$, 
and for the case of $H_0$, we have 
$|| H_{\rm pert} || \approx 20 \tau N_{\rm qubit} J\Omega$, 
where $N_{\rm qubit}$ is the number of connected qubits. 
As long as $N_{\rm qubit}$ is not large, 
these perturbation terms can be neglected 
by repeating Eq.(\ref{ABAB}) with $J_{ij}t_0/n \ll 1$.
Hereafter, we consider the case of $n=1$ for simplicity.
Note that the procedure described in this section can be easily extended to three-dimensionally (3D) arrayed qubits.

\subsection{Estimation of elapsed time}\label{sec:time}
In order to estimate an operation time of pulse manipulations, 
we express the time for single-qubit rotation as $\tau_{\rm rot}$.
For preparing a single Hamiltonian $H_0$, 
it takes an extra time of $5\tau_{\rm rot}$, because,
in Eq.~(\ref{ABAB}), four Hamiltonians $A$, $B$, $B'$, and $A'$
are transformed from $H_q$ by being sandwiched by $\pi$-pulses.
It also takes extra times of $4\tau_{\rm rot}$  and  $5\tau_{\rm rot}$ to 
obtain  $\exp ( i \tau_{\rm op} H_{\rm op})$  and $\exp (-i \tau_{\rm op} H_{\rm op})$,
respectively, in Eq.(\ref{rho}).
In the latter case,  $\tau_{\rm rot}$ is required to reverse the sign of $H_{\rm op}$.
Thus, for $N_{\rm op}$ qubit-qubit operations, it takes a time of 
$N_{\rm op} [2\tau_{\rm op}+9\tau_{\rm rot}]$. 

In the following, we would like to address the feasibility of our scheme in a
typical superconducting qubit system. Note that our qubit lattice model can 
be applied not only to solid-state coupling qubits~\cite{Zagoskin,Grajcar,Niskanen,Ashhab}, 
but also to circuit-QED qubits~\cite{Schoelkopf,NTT,Houck,Paik}.
For two superconducting qubits in a circuit-QED setup 
the effective inter-qubit interaction can be treated as $XY$
type~\cite{Blais++:04,toffoli}.  For instance, for $g/\Delta=0.1$, $g/(2\pi)=200$
MHz, $\Delta/(2\pi)=2$ GHz, where $g$ is the Jaynes-Cummings coupling
constant and $\Delta$ the detuning between the resonator frequency and
the qubit splitting, we have $J/(2\pi)=20$ MHz. Thus, $\tau_{\rm op}\approx 6.25$~ns. 
We also take $\tau_{\rm rot}\approx
1$ ns~\cite{qecc-basel}. 
The criterion is whether all pulse sequences can be done during the dephasing time $T_2$.
We will show that all generation times are less than 300~ns. 
Thus, if we assume $T_2 \sim$ 10 to 20~$\mu$s with well-controlled pulses, 
which was realized by Paik {\it et al}~\cite{Paik},
we will be able to use the standard QECC process and correct qubit errors, 
as long as the number of errors is small.

\section{Generation of stabilizer code from conventional Hamiltonian}\label{sec:code}
Here, we show concrete pulse sequences to produce the target stabilizer Hamiltonians
of the three major codes: the nine-qubit code, the five-qubit code, and the Steane code. 
In general, it is difficult to find a pulse sequence of the transformation 
from the conventional two-body solid-state Hamiltonian to 
the target stabilizer Hamiltonian, because the target Hamiltonians have Pauli matrices whose form is complicated. 
The best way to look for an appropriate pulse sequence is to change the target stabilizer Hamiltonian 
into single-qubit Hamiltonian, 
because it is easier to reduce 
the number of multiplications of the Pauli matrices to single-qubit Hamiltonian. 
In the following, we show the transformation process of $H_{\rm stab}$ of the three major codes 
to the initial single-qubit Hamiltonian.
We also count the number of pulses and estimate generation time of the codes.  
We show that the direct generation of $H_{\rm stab}$ is more effective 
than the previous method~\cite{qecc-basel} in which $G_i$ is generated one by one.
The comparison of the present results with those of the previous results
is summarized in Tables I and II.

\subsection{Nine-qubit code}
We would like to start from Shor's nine-qubit code that was the first advanced QECC to be invented~\cite{Shor}.
This code can correct single-qubit error ($n=9$, $k=1$), and the number of 
stabilizers is $l=8$. The stabilizers are given by $G_1=Z_1Z_2$, $G_2=Z_2Z_3$, $G_3=Z_4Z_5$, $G_4=Z_5Z_6$
$G_5=Z_7Z_8$, $G_6=Z_8Z_9$ $G_7=X_1X_2X_3X_4X_5X_6$, and $G_8=X_4X_5X_6X_7X_8X_9$~\cite{Gottesman,Nielsen}.
Then, the target stabilizer Hamiltonian is given by 
$H^{\rm 9code}=\sum_{i=1}^{8}G_i$ in which $\Omega_i$ are omitted, 
and we treat $H^{\rm 9code}=\sum_{i=1}^{8}G_i$ instead of $H^{\rm 9code}=-\sum_{i=1}^{8}G_i$
for clarity. We will treat the stabilizer Hamiltonians of the five-qubit code and the Steane code similarly.
We consider how this target Hamiltonian is transformed to 
a single-qubit Hamiltonian by using Eqs.~(\ref{XYa1})-~(\ref{XYc1}) for the XY interaction 
or Eqs.~(\ref{ZZa1})-~(\ref{ZZc1}) for the Ising interaction. Let us first consider a case of the 
XY interaction. $H^{\rm 9code}$ is changed as follows:
\begin{widetext}
\begin{eqnarray}
H^{\rm 9code}&=& Z_1Z_2+Z_2Z_3+Z_4Z_5+Z_5Z_6 +Z_7Z_8+Z_8Z_9
    +X_1X_2X_3X_4X_5X_6+X_4X_5X_6X_7X_8X_9, \ \ 
: (x \leftrightarrow z:2,4,6,8),
 \nonumber \\
&\rightarrow & Z_1X_2+X_2Z_3+X_4Z_5+Z_5X_6 +Z_7X_8+X_8Z_9
    -X_1Z_2X_3Z_4X_5Z_6-Z_4X_5Z_6X_7Z_8X_9, \ \ 
\nonumber \\  
& & \hspace{10cm}: H_{XY}^{12}+H_{XY}^{34}+H_{XY}^{56}+H_{XY}^{78},
 \nonumber \\
&\rightarrow &
 -Y_1-Y_1Z_2Z_4-Y_3Z_4Z_6-Y_5 -Y_7-Y_7Z_8Z_9
 +Y_2Y_4Y_6-Z_3Y_6Y_8X_9, \ \ : (y\leftrightarrow z:1,5,7)(x\leftrightarrow z:9)
 \nonumber \\
&\rightarrow &
 -Z_1-Z_1Z_2Z_4+Y_3Z_4Z_6-Z_5 -Z_7-Z_7Z_8X_9
 +Y_2Y_4Y_6+Z_3Y_6Y_8Z_9, \ \ : H_{XY}^{34}+H_{XY}^{56}+H_{XY}^{89} 
 \nonumber \\
&\rightarrow &
 -Z_1-Z_1Z_2Z_3+X_4Z_5-Z_6 -Z_7+Z_7Y_8
 +Y_2X_3Z_4X_5Z_6+Z_4X_5Z_6X_9, \ \ : (x\leftrightarrow z:3,4,5,9)
 \nonumber \\
&\rightarrow &
 -Z_1-Z_1Z_2X_3-Z_4X_5-Z_6 -Z_7+Z_7Y_8
 +Y_2Z_3X_4Z_5Z_6+X_4Z_5Z_6Z_9, \ \ : H_{XY}^{23}+H_{XY}^{45}+H_{XY}^{78} 
 \nonumber \\
&\rightarrow &
 -Z_1+Z_1Y_2+Y_4-Z_6 -Z_8+X_7
 -X_3Y_5Z_6-Y_5Z_6Z_9, \ \ : (y\leftrightarrow z:4)(x\leftrightarrow z:7), \ H_{XY}^{12}+H_{XY}^{56}+H_{XY}^{89} 
 \nonumber \\
&\rightarrow &
 -Z_2+X_1+Z_4-Z_5 -Z_9-Z_7
 -X_3X_6-X_6Z_8, \ \ : (x\leftrightarrow z:1,3) 
 \nonumber \\
&\rightarrow &
 -Z_2-Z_1+Z_4-Z_5 -Z_9-Z_7
 +Z_3X_6-X_6Z_8, \ \ : H_{XY}^{34}+H_{XY}^{78} 
 \nonumber \\
&\rightarrow &
 -Z_2-Z_1+Z_3-Z_5 -Z_9-Z_8
 +Z_4X_6-X_6Z_7, \ \ : H_{XY}^{45}+H_{XY}^{67} 
 \nonumber \\
&\rightarrow &
 -Z_2-Z_1+Z_3-Z_4 -Z_9-Z_8
 -Z_5Z_6Y_7+Y_7, \ \ : (x\leftrightarrow z:5)(y\leftrightarrow z:7)
 \nonumber \\
&\rightarrow &
 -Z_2-Z_1+Z_3-Z_4 -Z_9-Z_8
 -X_5Z_6Z_7+Z_7, \ \ : H_{XY}^{56}  
 \nonumber \\
&\rightarrow &
 -Z_2-Z_1+Z_3-Z_4 -Z_9-Z_8
 +Y_6Z_7+Z_7, \ \ : H_{XY}^{67} 
 \nonumber \\
&\rightarrow &
 -Z_2-Z_1+Z_3-Z_4 -Z_9-Z_8
 +X_7+Z_6,
\label{9codeXY} 
\end{eqnarray}
Applied pulses are shown after the colon in each line.
Regarding the notation, the $H_{XY}^{ij}$ on the right-hand side of each line 
after the colon shows that we apply Eq.~(\ref{rho}) to the Hamiltonian of the left-hand side.
For example, the second line of the above equation means that
\begin{equation}
e^{ i\frac{\pi}{4}[H_{XY}^{12}+H_{XY}^{34}+H_{XY}^{56}+H_{XY}^{78}]}  H^{\rm 9code}
e^{ -i\frac{\pi}{4}[H_{XY}^{12}+H_{XY}^{34}+H_{XY}^{56}+H_{XY}^{78}]}.
\end{equation}
The notation such as $(y\leftrightarrow z:1,5,7,9)$ shows that single-qubit $\pi$-rotation
is applied to qubits 1,5,7 and 9 around the $x$-axis. 
Thus when we start an initial Hamiltonian given by
\begin{equation}
H_{\rm ini}^{\rm 9code}=\Omega_1 X_1+\Omega_2 X_2+\Omega_3 X_3+\Omega_4 X_4+\Omega_6 X_6
+\Omega_7 X_7+ \Omega_8 X_8+ \Omega_9 X_9,
\label{9codeini}
\end{equation}
we can produce the stabilizer Hamiltonian $H^{\rm 9code}$ by using the pulse sequence 
described by the reverse operations of Eq.(\ref{9codeXY}).
The initial Hamiltonian Eq.(\ref{9codeini}) is obtained by 
$e^{-i t H_0} e^{i \pi X_5/2 }e^{-i t H_0} e^{-i \pi X_5/2 }$ in which
$e^{-i t H_0}$ term is obtained from $H_q$ as shown in the previous section.
For the Ising interaction, we obtain
\begin{eqnarray}
H^{\rm 9code} &\rightarrow & 
 X_1Z_2+Z_2Z_3+X_4Z_5+Z_5X_6 +Z_7Z_8+Z_8X_9
    -Z_1X_2X_3Z_4X_5Z_6-Z_4X_5Z_6X_7X_8Z_9, \ \ 
\nonumber \\  
& & \hspace{10cm}: H_{\rm Ising}^{12}+H_{\rm Ising}^{56}+H_{\rm Ising}^{89}
 \nonumber \\
&\rightarrow &
 Y_1+Z_2Z_3+X_4Z_5+Y_6 +Z_7Z_8+Y_9
    -Y_2X_3Z_4Y_5-Z_4Y_5X_7Y_8, \ \ : (y\leftrightarrow z:2,8)
 \nonumber \\
&\rightarrow &
 Y_1-Y_2Z_3+X_4Z_5+Y_6 -Z_7Y_8+Y_9
    -Z_2X_3Z_4Y_5-Z_4Y_5X_7Z_8, \ \ : H_{\rm Ising}^{23}+H_{\rm Ising}^{45}+H_{\rm Ising}^{78}
 \nonumber \\
&\rightarrow &
 Y_1+X_2+Y_4+Y_6 +X_8+Y_9
    +Y_3X_5+X_5Y_7, \ \ : H_{\rm Ising}^{34}+H_{\rm Ising}^{67}
 \nonumber \\
&\rightarrow &
 Y_1+X_2-Z_3X_4-X_6Z_7 +X_8+Y_9
    -X_3Z_4X_5-X_5Z_6X_7, \ \ :(x\leftrightarrow z:3,4,5,6,7)
 \nonumber \\
&\rightarrow &
 Y_1+X_2+X_3Z_4+Z_6X_7 +X_8+Y_9
    -Z_3X_4Z_5-Z_5X_6Z_7, \ \ : H_{\rm Ising}^{45}+H_{\rm Ising}^{67}
 \nonumber \\
&\rightarrow &
 Y_1+X_2+X_3Z_4+Y_7 +X_8+Y_9
    -Z_3Y_4-Z_5Y_6, \ \ : H_{\rm Ising}^{34}+H_{\rm Ising}^{56}
 \nonumber \\
&\rightarrow &
 Y_1+X_2+Y_3+Y_7 +X_8+Y_9
    +X_4+X_6, \ \ 
    \label{9codeZZ} 
\end{eqnarray}
\end{widetext}
After single-qubit rotations, we obtain an initial Hamiltonian:
\begin{eqnarray}
H_{\rm ini}^{\rm 9code}&=& \Omega_1 X_1+\Omega_2 X_2+\Omega_3 X_3+\Omega_4 X_4
\nonumber \\
&+&\Omega_6 X_6 +\Omega_7 X_7 +\Omega_8 X_8+\Omega_9 X_9.
\end{eqnarray}
This Hamiltonian is obtained by eliminating $X_5$ term in $H_0$ as in the case of the XY interaction.

Let us count the number of pulses necessary to obtain the nine-qubit code.
Because the present method mainly relies on the control of many pulses, 
as the number of pulses increases, pulse errors become the principal origin of 
decoherence. Thus, the number of pulses is an indicator 
of decoherence in which it is desirable to have fewer pulses.  
Eq.(\ref{9codeini}) shows that eight qubit-qubit interaction processes and five single-qubit 
rotation processes are needed. 
Note that because $\tau_{\rm op} > \tau_{\rm rot}$, the sixth operations of Eq.(\ref{9codeXY}) 
can be represented by $\tau_{\rm rot}$. 
From the result of Sec.III, it takes a time of $N_{\rm op} [2\tau_{\rm op}+9\tau_{\rm rot}]$ 
for $N_{\rm op}$ uses of $H_{XY}^{ij}$. The initial state Eq.(\ref{9codeini}) is 
obtained by twice using the generating process of $H_0$, and thus it takes a time of $10\tau_{\rm rot}$.
Thus, we need a time of
\begin{equation}
\tau^{\rm 9code(new)}_{XY}=8[2\tau_{\rm op}+9 \tau_{\rm rot}]+12\tau_{\rm rot}+10\tau_{\rm rot}=
16\tau_{\rm op}+94 \tau_{\rm rot}. 
\end{equation} 
In order to compare the present method with that of Ref.~\cite{qecc-basel},
let us consider constructing $H_{\rm stab}$ of the XY interaction by 
summing up $G_i$ as in Ref.~\cite{qecc-basel}.
For $G_1 \sim G_6$, it takes a time of  
$[2\tau_{\rm op}+9\tau_{\rm rot}] + (4+10)\tau_{\rm rot}
= 2\tau_{\rm op}+23 \tau_{\rm rot}$, 
because $G_i \rightarrow Z_iX_{i+1} \rightarrow Y_i \rightarrow X_i$.
For $G_7$, we have
\begin{eqnarray}
G_7 
&= & X_1X_2X_3X_4X_5X_6, \ \ : (x \leftrightarrow z:2,5)
\nonumber \\
&\rightarrow & X_1Z_2X_3X_4Z_5X_6, \ \ : H_{XY}^{12}+H_{XY}^{56}
\nonumber \\
&\rightarrow & Y_2X_3X_4Y_5, \ \ : (x\leftrightarrow z:3,4)
\nonumber \\
&\rightarrow & Y_2Z_3Z_4Y_5, \ \ : H_{XY}^{23}+H_{XY}^{45}
\nonumber \\
&\rightarrow & X_3X_4, \ \ : (x\leftrightarrow z:3)
\nonumber \\
&\rightarrow & -Z_3X_4, \ \ : H_{XY}^{34}
\nonumber \\
&\rightarrow & Y_3, \ \ : (y\leftrightarrow x:3)
\nonumber \\
&\rightarrow & X_3.
\end{eqnarray}
Thus, it takes a time of
$ 3[2\tau_{\rm op}+9\tau_{\rm rot}]+8\tau_{\rm rot}+10\tau_{\rm rot}=6\tau_{\rm op}+45\tau_{\rm rot}$
for obtaining $G_7$ and $G_8$. Thus, total generation time for the nine-qubit code 
by the method of Ref.~\cite{qecc-basel} is given by $\tau^{\rm 9code(old)}_{XY}=24\tau_{\rm op}+228\tau_{\rm rot}$. 
Thus, the number of the qubit-qubit interaction of the present method is reduced to two-thirds of that of the previous method and
the number of the single-qubit rotations is reduced to 41.2 \% of that of the previous method. 
When we use the experimental values in Sec.\ref{sec:time},
$\tau^{\rm 9code(new)}_{XY}= 194$~ns and $\tau^{\rm 9code(old)}_{XY}=376$~ns, thus 48.7\% reduction 
of the operation time is achieved.
For the Ising interaction, from Eq.~(\ref{9codeZZ}), we obtain a time of the operation given by
\begin{equation}
\tau^{\rm 9code(new)}_{\rm Ising}=10\tau_{\rm op}+63 \tau_{\rm rot}=125.5 {\rm ns}.
\end{equation} 
Here, we used the experimental value of $\tau_{\rm op}\approx$ 6.25~ns and $\tau_{\rm rot}\approx$ 1~ns 
(See Sec.~\ref{sec:time}).
$G_1 \sim G_6$ have the form of the two-body interaction, thus they are directly extracted from $H_q$ 
as shown in Sec.\ref{sec:extract}. 
Thus it takes a time of $4\tau_{\rm rot}$ for each process. 
$G_7$ and $G_8$ are reduced to $Z_3Z_4$ and $Z_6Z_7$ with a time of 
$4\tau_{\rm op}+28\tau_{\rm rot}$, respectively. Therefore,
we obtain 
$\tau^{\rm 9code(old)}_{\rm Ising}=8\tau_{\rm op}+80 \tau_{\rm rot}=130 {\rm ns}$.
For this case, 3.5~\% reduction of time is achieved.

\subsection{Five-qubit code}
Next, we consider $H_{\rm stab}$ of the five-qubit code ($n=5$ and $k=1$). 
The stabilizers $G_i (i=1,..,4)$ of this code are given by
$G_1 = X_1 Z_2Z_3 X_4$, 
$G_2 = X_2 Z_3 Z_4 X_5$,
$G_3 = X_3 Z_4 Z_5 X_1$, and
$G_4 = X_4 Z_5 Z_1 X_2$~\cite{Gottesman,Nielsen}.  
The process of constructing $H^{\rm 5code}\equiv \sum_{i=1}^4 G_i$ is 
obtained by changing $H^{\rm 5code}$ 
reversely into a single-qubit Hamiltonian. 
For $XY$ model, this process is obtained by
\begin{widetext}
\begin{eqnarray}
H^{\rm 5code}&=& 
 X_1Z_2Z_3X_4 +X_2Z_3Z_4X_5+X_3Z_4Z_5X_1+X_4Z_5Z_1X_2, \ :  H_{XY}^{12}+H_{XY}^{34}
\nonumber \\
&\rightarrow &
 Y_2Y_3 -Y_1Z_2Z_3Z_4X_5+Y_4Z_5Z_1Y_2+Y_3Z_4Z_5Y_1, \ :  (y\leftrightarrow z:2)
\nonumber \\
&\rightarrow &
 Z_2Y_3 +Y_1Y_2Z_3Z_4X_5+Y_4Z_5Z_1Z_2+Y_3Z_4Z_5Y_1, \ :  H_{XY}^{23}+H_{XY}^{45}
\nonumber \\
&\rightarrow &
 X_2 -Y_1X_3Y_4+Z_1Z_3X_5+Y_1X_2Z_3Z_4Z_5, \ : (x\leftrightarrow z:2,3)
\nonumber \\
&\rightarrow &
-Z_2 +Y_1Z_3Y_4+Z_1X_3X_5-Y_1Z_2X_3Z_4Z_5, \ :  H_{XY}^{12}+H_{XY}^{34}
\nonumber \\
&\rightarrow &
-Z_1 +Z_1X_2X_3-Z_2Z_3Y_4X_5+X_2Y_4Z_5, \ :  (x\leftrightarrow z:2)(y\leftrightarrow z:4,5)
\nonumber \\
&\rightarrow &
-Z_1 -Z_1Z_2X_3-X_2Z_3Z_4X_5+Z_2Z_4Y_5, \ :  H_{XY}^{23}+H_{XY}^{45}
\nonumber \\
&\rightarrow &
-Z_1 +Z_1Y_2-Y_3Y_4+Z_3X_4, \ : (y\leftrightarrow z:4)
\nonumber \\
&\rightarrow &
-Z_1 +Z_1Y_2-Y_3Z_4+Z_3X_4, \ :H_{XY}^{12}+H_{XY}^{34}
\nonumber \\
&\rightarrow &
-Z_2 +X_1-X_4-Y_3,
\label{5codeXY}
\end{eqnarray}
Thus, the initial Hamiltonian is given by
\begin{equation}
H_{\rm ini}^{\rm 5code}=\Omega_1 X_1+\Omega_2 X_2 +\Omega_3 X_3+\Omega_4 X_4.
\label{ini_five_xy}
\end{equation}
The time of constructing this code is given by
\begin{equation}
\tau^{\rm 5code(new)}_{XY}=5[2\tau_{\rm op}+9\tau_{\rm rot}]+10\tau_{\rm rot}+10\tau_{\rm rot}
=10\tau_{\rm op}+65\tau_{\rm rot}=127.5 {\rm ns}.
\label{time5code}
\end{equation} 
If we use the previous method in Ref.~\cite{qecc-basel}, we have
$
\tau^{\rm 5code(old)}_{XY}=24 \tau_{\rm op}+162\tau_{\rm rot}=312 {\rm ns}.
$
This result is a little different from that in Ref.~\cite{qecc-basel}
in that here we start from 2D Hamiltonian.
Thus, 59.1 \% reduction of time is expected with the present method.
For the Ising interaction, we have
\begin{eqnarray}
H^{\rm 5code}&\rightarrow & 
 Z_1X_2X_3Z_4 +Z_2X_3X_4Z_5+Z_1Z_3X_4X_5+X_1Z_2Z_4X_5, \ :  H_{\rm Ising}^{23}
\nonumber \\
&\rightarrow &
 Z_1X_2X_3Z_4 +Y_3X_4Z_5+Z_1Z_3X_4X_5+X_1Z_2Z_4X_5, \ :  (x\leftrightarrow z:1,2)(y\leftrightarrow z:3)
\nonumber \\
&\rightarrow &
-X_1Z_2X_3Z_4 +Z_3X_4Z_5-X_1Y_3X_4X_5-Z_1X_2Z_4X_5, \ :  H_{\rm Ising}^{12}+H_{\rm Ising}^{34}
\nonumber \\
&\rightarrow &
-Y_1Y_3 +Y_4Z_5-Y_1Z_2Y_3X_4X_5-Y_2Z_4X_5, \ :   (y\leftrightarrow z:1,2,4)(x\leftrightarrow z:5)
\nonumber \\
&\rightarrow &
-Z_1Y_3 +Z_4X_5-Z_1Y_2Y_3X_4Z_5-Z_2Y_4Z_5, \ :   H_{\rm Ising}^{12}+H_{\rm Ising}^{45}
\nonumber \\
&\rightarrow &
-Z_1Y_3 +Y_5+X_2Y_3Y_4+Z_2X_4, \ :   (x\leftrightarrow z:2)
\nonumber \\
&\rightarrow &
-Z_1Y_3 +Y_5-Z_2Y_3Y_4+X_2X_4, \ :   H_{\rm Ising}^{23}
\nonumber \\
&\rightarrow &
 Z_1Z_2X_3 +Y_5+X_3Y_4+Y_2Z_3X_4, \ :   (y\leftrightarrow z:2) (x\leftrightarrow z:3,4)
\nonumber \\
&\rightarrow &
 Z_1Y_2Z_3 +Y_5-Z_3Y_4-Z_2X_3Z_4, \ :  H_{\rm Ising}^{23}
\nonumber \\
&\rightarrow &
-Z_1X_2 +Y_5-Z_3Y_4-Y_3Z_4, \ :  H_{\rm Ising}^{12}+H_{\rm Ising}^{34}
\nonumber \\
&\rightarrow &
-Y_2 +Y_5+X_4+X_3, 
\end{eqnarray}
\end{widetext}
Thus, the initial Hamiltonian from which $H_{\rm stab}$ is derived is given by 
\begin{equation}
H_{\rm ini}^{\rm 5code}= \Omega_2 X_2+\Omega_3 X_3+\Omega_4 X_4 +\Omega_5 X_5, 
\end{equation}
The time for the generation of this code is 
$6[2\tau_{\rm op}+9\tau_{\rm rot}]+12\tau_{\rm rot}+10\tau_{\rm rot}
=151
$~ns.
Because 
$G_1=X_1Z_2Z_3X_4 \rightarrow Z_1X_2X_3Z_4\rightarrow Y_2Y_3 \rightarrow Z_2Z_3$, 
it takes a time of 
$[2\tau_{\rm op}+9\tau_{\rm rot}]+4\tau_{\rm rot}+4\tau_{\rm rot}
=2\tau_{\rm op}+17\tau_{\rm rot}$  to obtain $G_1$ and $G_2$. $G_3$ is estimated 
from 
\begin{eqnarray}
G_3
&= & X_1X_3Z_4Z_5, \ \ : (x \leftrightarrow z:1,4)
\nonumber \\
&\rightarrow & -Z_1X_3X_4Z_5, \ \ : H_{XY}^{23}+H_{XY}^{45}
\nonumber \\
&\rightarrow & -Z_1Z_2Y_3Y_4, \ \ : (y\leftrightarrow z:2,4)
\nonumber \\
&\rightarrow &  Z_1Y_2Y_3Z_4, \ \ : H_{XY}^{12}+H_{XY}^{34}
\nonumber \\
&\rightarrow & X_2X_3, \ \ : (x\leftrightarrow z:2,3)
\nonumber \\
&\rightarrow & Z_2Z_3.
\end{eqnarray}
Thus it takes $4\tau_{\rm op}+28\tau_{\rm rot}$. Similarly, it takes 
$6\tau_{\rm op}+35\tau_{\rm rot}$ for $G_4$. Therefore, in total, 
it takes $14\tau_{\rm op}+97\tau_{\rm rot}=184.5$~ns for summing up $G_1 \sim G_4$ in 
the Ising interaction. In this case the present method reduces the generation 
time by 18.2\%.

\subsection{Steane code}
The stabilizers of the Steane code are described by 
$G_1 = X_1 X_2 X_3 X_4$,
$G_2 = X_1 X_2 X_5 X_6$,  
$G_3 = X_1 X_3 X_5 X_7$, 
$G_4 = Z_1 Z_2 Z_3 Z_4$, 
$G_5 = Z_1 Z_2 Z_5 Z_6$,  and
$G_6 = Z_1 Z_3 Z_5 Z_7$~\cite{Gottesman,Nielsen}. 
Because $G_4$, $G_5$ and $G_7$ are obtained from $G_1$, $G_2$ and $G_3$ by applying $\pi$-pulses,
we first consider the generation process of $H_X^{Steane}\equiv G_1+G_2+G_3$.
The process of the construction of $H_X^{\rm Steane}$ is obtained by resolving it to a single-qubit Hamiltonian.
For the case of $XY$ Hamiltonian, this process is given by,
\begin{widetext}
\begin{eqnarray}
H_X^{\rm Steane}&=& 
 X_1X_2X_3X_4 +X_1X_2X_5X_6+X_1X_3X_5X_7, \ :  (x\leftrightarrow z :2,3,5)\nonumber \\
&\rightarrow &
 X_1Z_2Z_3X_4 +X_1Z_2Z_5X_6+X_1Z_3Z_5X_7, \ :  H_{XY}^{12}+H_{XY}^{34}+H_{XY}^{56}\nonumber \\
&\rightarrow &
 Y_2Y_3 +Y_2Y_5-Z_1Y_2Z_4Z_6X_7, \ : (y\leftrightarrow z:2,5) \nonumber \\
&\rightarrow &
 Z_2Y_3 +Z_2Z_5-Z_1Z_2Z_4Z_6X_7, \ : H_{XY}^{23}+H_{XY}^{45}+H_{XY}^{67} \nonumber \\
&\rightarrow &
 X_2 +Z_3Z_4+Z_1Z_3Z_5Y_6, \ : (x\leftrightarrow z:2,4) \nonumber \\
&\rightarrow &
-Z_2 +Z_3X_4+Z_1Z_3Z_5Y_6, \ : H_{XY}^{12}+H_{XY}^{34}+H_{XY}^{56} \nonumber \\
&\rightarrow &
-Z_1 -Y_3+Z_2Z_4X_5, \ : (y \leftrightarrow z:3) \nonumber \\
&\rightarrow &
-Z_1 -Z_3+Z_2Z_4X_5, \ : H_{XY}^{23}+H_{XY}^{45}  \nonumber \\
&\rightarrow &
-Z_1 -Z_2-Z_3Y_4, \ : H_{XY}^{34}  \nonumber \\
&\rightarrow &
-Z_1 -Z_2-X_3. \ 
\label{CSSXY}
\end{eqnarray}
\end{widetext}
Thus, we obtain the initial Hamiltonian: 
\begin{equation}
H_{X:{\rm ini}}^{\rm Steane}= \Omega_1 X_1+\Omega_2 X_2+\Omega_3 X_3, 
\end{equation}
The time of generation of $H_{\rm X}^{\rm Steane}$ is the same as Eq.(\ref{time5code}).
For the previous method, the time for obtaining $H_{\rm X}^{\rm Steane}$ 
is given by $22\tau_{\rm op}+143\tau_{\rm rot}$. 
As mentioned above, because $H_Z^{\rm Steane}\equiv G_4+G_5+G_6$ is obtained by
\begin{equation}
H_Z^{\rm Steane}=e^{-i\pi (Y_1+Y_2+Y_3)/4}H_X^{\rm Steane}e^{i\pi (Y_1+Y_2+Y_3)/4},
\end{equation}
the generation time of $H_Z^{\rm Steane}$ is increased by $2\tau_{\rm rot}$
compared with that of $H_X^{\rm Steane}$. Then, the time of obtaining the 
Steane code by the present method is given by
\begin{equation}
\tau^{\rm Steane(new)}_{XY}
= 20 \tau_{\rm op}+132\tau_{\rm rot} =257 {\rm ns}.
\label{timeSteane}
\end{equation} 
When we use the previous method in Ref.~\cite{qecc-basel}, the time for the code 
generation is given by $\tau^{\rm Steane(old)}_{XY}=44\tau_{\rm op}+288\tau_{\rm rot}=563$~ns.
Thus, 54.4\% reduction of time is expected. 
\begin{widetext}
For the Ising Hamiltonian, we have
\begin{eqnarray}
H_X^{\rm Steane}&=& 
 X_1X_2X_3X_4 +X_1X_2X_5X_6+X_1X_3X_5X_7, \ :  (z \leftrightarrow x:1,4,6) \nonumber \\
&\rightarrow &
 Z_1X_2X_3Z_4 +Z_1X_2X_5Z_6-Z_1X_3X_5X_7, \ :  H_{\rm Ising}^{12}+H_{\rm Ising}^{34}+H_{\rm Ising}^{56}
\nonumber \\
&\rightarrow &
 Y_2Y_3 +Y_2Y_5-Z_1Y_3Z_4Y_5Z_6X_7, \ :  (y\leftrightarrow z:3)(x\leftrightarrow z:6,7)
\nonumber \\
&\rightarrow &
 Y_2Z_3 +Y_2Y_5+Z_1Z_3Z_4Y_5X_6Z_7, \ :  H_{\rm Ising}^{23}+H_{\rm Ising}^{45}+H_{\rm Ising}^{67}
\nonumber \\
&\rightarrow &
-X_2 +X_2Z_3Z_4X_5-Z_1Z_3X_5Y_6, \ :  (x\leftrightarrow z:2,3,4,5)
\nonumber \\
&\rightarrow &
 Z_2 +Z_2X_3X_4Z_5+Z_1X_3Z_5Y_6, \ :  H_{\rm Ising}^{23}+H_{\rm Ising}^{45}
\nonumber \\
&\rightarrow &
 Z_2 +Y_3Y_4+Z_1Z_2Y_3Z_5Y_6, \ :  (y\leftrightarrow z:2,4,5,6)
\nonumber \\
&\rightarrow &
-Y_2 +Y_3Z_4+Z_1Y_2Y_3Y_5Z_6, \ :  H_{\rm Ising}^{12}+H_{\rm Ising}^{34}+H_{\rm Ising}^{56}
\nonumber \\
&\rightarrow &
 Z_1X_2 -X_3-X_2X_3Z_4X_5, \ :  (x\leftrightarrow z:2,4,5)
\nonumber \\
&\rightarrow &
-Z_1Z_2 -X_3-Z_2X_3X_4Z_5, \ :  H_{\rm Ising}^{23}+H_{\rm Ising}^{45}
\nonumber \\
&\rightarrow &
-Z_1Z_2 -Z_2Y_3-Y_3Y_4,
\label{CSSIsing}
\end{eqnarray}

\end{widetext}
Thus, the initial Hamiltonian is given by 
\begin{equation}
H_{X:{\rm ini}}^{\rm Steane}= J_{12}Z_1Z_2 +J_{23}Z_2Z_3+J_{34}Z_3Z_4, 
\label{SteaneIsingIni}
\end{equation}
This Hamiltonian is obtained by erasing $H_{\rm Ising}^{45}$ from $H_{\rm Ising}$. 
$H_{\rm Ising}$ is obtained by applying $\pi$-pulses to all qubits in $B$ of Eq.~(\ref{AB}). 
Then, Eq.(\ref{SteaneIsingIni}) is obtained by applying a $\pi$-pulse 
only to qubit 5 in $B$ of Eq.~(\ref{AB}) for $H_{\rm Ising}$. Then, the time of 
the preparation of Eq.~(\ref{SteaneIsingIni}) is estimated by 4$\tau_{\rm rot}$.
Therefore, the total time of the generation of Eq.~(\ref{CSSIsing}) is given by
$5[2\tau_{\rm op}+9\tau_{\rm rot}]+12\tau_{\rm rot}+4\tau_{\rm rot}
=123.5$~ns, and 
$\tau^{\rm Steane(new)}_{\rm Ising}
=20\tau_{\rm op}+124\tau_{\rm rot}=249$~ns.
On the other hand, when we use the previous method, 
times for generating $G_1$, $G_2$ and $G_3$ are 
$4\tau_{\rm op}+26\tau_{\rm rot}$, 
$6\tau_{\rm op}+35\tau_{\rm rot}$, and
$8\tau_{\rm op}+48\tau_{\rm rot}$, respectively. Therefore, we obtain
$\tau^{\rm Steane(old)}_{\rm Ising}=36\tau_{\rm op}+220\tau_{\rm rot}=445$~ns, 
resulting in 44\% reduction of time.

All the results of the above-mentioned three codes 
are summarized in Tables I and II for the $XY$ interaction and Ising interaction, respectively. 
From Tables I and II, we can see the large reduction of 
the generation time is achieved in the $XY$ interaction.

\begin{table*}
\begin{tabular}{l|l|l|l|l|c}
\hline\hline
$XY$ 
 & \multicolumn{2}{c|}{Previous generation time}
 & \multicolumn{2}{c|}{New generation time }
 & Improvement \\
\hline
Nine-qubit code  
& $24\tau_{\rm op}+228\tau_{\rm rot}$ & 378~ns
& $16\tau_{\rm op}+92\tau_{\rm rot}$ & 194~ns 
& 48.7~\% \\

Five-qubit code   
& $24\tau_{\rm op}+162\tau_{\rm rot}$ & 312~ns
& $10\tau_{\rm op}+65\tau_{\rm rot}$  & 127.5~ns
& 59.1~\% \\

Steane code  
&$44\tau_{\rm op}+288\tau_{\rm rot}$ & 563~ns
&$20\tau_{\rm op}+132\tau_{\rm rot}$ & 257~ns 
& 54.4~\% \\
\hline\hline
\end{tabular}
\begin{flushleft}
TABLE I. The generation time of the stabilizer Hamiltonian of the
$XY$ interaction. ``New generation time" is a generation time 
of the stabilizer Hamiltonian by using the proposed method. 
``Previous generation time" is a time, 
estimated by using the previous method~\cite{qecc-basel}.
$\tau_{\rm op} = \pi/(4 J)$.
$\tau_{\rm rot}$ represents a time of a single qubit rotation.
We take $\tau_{\rm op}=6.25$~ns and $\tau_{\rm rot}=1$~ns 
(Sec.~\ref{sec:time}).
``Improvement" is a ratio of reduction of time of the new generation, 
calculated from the 3rd and 5th columns.
\end{flushleft}
\end{table*}
\begin{table*}
\begin{tabular}{l|l|l|l|l|c}
\hline\hline
Ising 
 & \multicolumn{2}{c|}{Previous generation time}
 & \multicolumn{2}{c|}{New generation time}
 & Improvement \\
\hline
Nine-qubit code  
& $8\tau_{\rm op}+80\tau_{\rm rot}$  & 130~ns
& $10\tau_{\rm op}+61\tau_{\rm rot}$ & 125.5~ns
& 3.5~\% \\

Five-qubit code   
& $14\tau_{\rm op}+97\tau_{\rm rot}$ & 184.5~ns 
& $12\tau_{\rm op}+76\tau_{\rm rot}$ & 151~ns
& 18.2~\% \\

Steane code  
&$36\tau_{\rm op}+220\tau_{\rm rot}$ & 445~ns
&$20\tau_{\rm op}+124\tau_{\rm rot}$ & 249~ns
&44~\% \\

\hline\hline
\end{tabular}
\begin{flushleft}
TABLE II. The generation time of the stabilizer Hamiltonian from the
Ising model. Parameters are the same as those in Table I.
\end{flushleft}
\end{table*}

\section{Creation of the standard codes}\label{sec:initial}
As briefly reviewed in Sec.~\ref{sec:review}, encoded states are generated by 
repeating measurements of  the stabilizers
$G_i$ ($i=1,..,l)$ for an initial state $\Pi_{i=1}^k|0\rangle_n$~\cite{Gottesman,Nielsen}. 
Considering that measurements induce extra decoherence, the effectiveness of this conventional method 
is limited. In Ref.~\cite{qecc-basel}, we presented the more effective method 
of directly generating logical states:
For any given code,
only those $G_j$ with $1 \le j \le m$ and $m \le n - k$ that contain
$X$ or $Y$ operators are needed for the preparation:
\begin{eqnarray}\label{eqn:CodeGen}
|\bar{c}_1...\bar{c}_k\rangle &=& (1+G_1)\cdots
(1+G_{m})\bar{X}^{c_1}_1
\cdots\bar{X}^{c_k}_k |0...0\rangle \nonumber \\
&= & \prod_{i=1}^{k} \bar{X}^{c_i}_i \prod_{j=1}^{m} 
\exp\left( -i \frac{\pi}{4}  \tilde{G}_j^{a_j} \right) |0...0\rangle\:,
\label{gen}
\end{eqnarray}
where $c_i=0,1$ and operators $\bar{X}_i$ act in the logical state space
$\{|\bar{0}\rangle_i$, $|\bar{1}\rangle_i\}$.  Here, $\tilde{G}_j^{a_j}$ denotes 
a modified stabilizer operator obtained from $G_j$
by replacing the $X$ operator acting on qubit $ a_j$ by a $Y$
operator, or vice versa. This is done in order to match the effect of an \emph{individual} 
factor $\exp [i(\pi/4) \tilde{G}_j^{a_j} ]$ with
the action of the projector $(1+G_j)$ when qubit $ a_j$ is in state
$|0\rangle$. To fulfill Eq.~\eqref{eqn:CodeGen} for all $1 \le j \le m$ \emph{simultaneously}, all
the $a_j$ have to be different and the modified stabilizers have to be generated
in an order such that prior to $\tilde{G}_j^{a_j}$ none of the $\tilde{G}_k^{a_k}$
with $k < j$ have acted on qubit $a_j$ with an $X$ or $Y$.
The time for generating the encoded state is given by
$\tau_{\rm stab}+(\sum_i c_i)\tau_{\rm rot}$.

Here, we extend this idea further and consider 
whether we can replace this equation by
\begin{eqnarray}
|\bar{0} \rangle &=& \exp \left( -i\frac{\pi}{4} \tilde{H}_{\rm stab} \right)|0\rangle, \\
\tilde{H}_{\rm stab} &\equiv & \sum_i \tilde{G}_i.
\label{gennew}
\end{eqnarray}
For the five-qubit code, we need
$\tilde{G}_1=Y_1Z_2Z_3X_4$, $\tilde{G}_2=X_2Z_3Z_4Y_5$, $\tilde{G}_3=X_1Y_3Z_4Z_5$, 
$\tilde{G}_4=Z_1 Y_2 X_4 Z_5$, and the multiplication is
carried out in the following order: 
$\exp[ i (\pi/4) \tilde{G}_2 ]$ $\exp[i (\pi/4) \tilde{G}_4 ]$
$\exp[ i (\pi/4) \tilde{G}_3 ]$ $\exp[i (\pi/4) \tilde{G}_1 ]$.
However, only $\tilde{G}_3$ and $\tilde{G}_4$ commute,
Thus, we cannot replace Eq.~(\ref{gen}) by Eq.~(\ref{gennew}).

For the Steane code, we need three generators:
\begin{eqnarray}
\tilde{G}_1&=&X_1X_2X_3Y_4, \\ 
\tilde{G}_2&=&X_1X_2X_5Y_6, \\
\tilde{G}_3&=&X_1X_3X_5Y_7.
\end{eqnarray}
Because these three generators mutually commute, such 
as $[\tilde{G}_i,\tilde{G}_j]=0$. Therefore 
we can apply Eq.~(\ref{gennew}) and reduce the generation time of the 
encoded state.
Thus, it is observed that sparse distribution of the Pauli 
operators in a logical qubit is preferable for the code generation,
because it results in simpler generation of encoded states. 

Next, we consider an encoding of unknown state $a|0\rangle +b|1\rangle$ to
$a|\bar{0}\rangle +b|\bar{1} \rangle $ ($a$ and $b$ are arbitrary complex numbers). 
Because, in Eq.(\ref{gen}), $\tilde{G}_j^{a_j}$ was 
introduced to hold 
$\exp[ -i (\pi/4)  \tilde{G}_j^{a_j} ]|0\rangle=(1+G_j )|0\rangle$, 
we need different operations for obtaining $|\bar{1} \rangle$. 
For simplicity, we consider $|\bar{1}\rangle=\bar{X}|\bar{0} \rangle$. 
Then, we can solve this problem if we can prepare a modified initial 
state for $|1\rangle$ defined by 
\begin{equation}
|\bar{1}\rangle' =\bar{M}^{-1} \bar{X} \bar{M} |0,...,0\rangle,
\end{equation}
with 
$\bar{M}\equiv \prod_{j=1}^{m} 
\exp [ -i (\pi/4)  \tilde{G}_j^{a_j}] $.
This is because we can use the following relation: 
\begin{eqnarray}
\bar{M} (a|0...0\rangle 
+ b  |\bar{1}\rangle') 
=a|\bar{0}\rangle +b|\bar{1}\rangle.
\end{eqnarray}
For the five-qubit code, $\bar{X}$ is given by $\bar{X}=X_1X_2X_3X_4X_5$~\cite{Gottesman},
and the modified initial state $|\bar{1}\rangle'$ is expressed by 
$|\bar{1}\rangle'=-\tilde{G}_3\tilde{G}_2 \bar{X} |00000\rangle=-|00010\rangle$.
This means that we can obtain an encoded unknown state $a|\bar{0}\rangle +b|\bar{1}\rangle$ 
when we encode an initial unknown state $a|0\rangle+b|1\rangle$ into 
the fourth qubit described by 
$|0\rangle_1|0\rangle_2|0\rangle_3(a|0\rangle_4-b|1\rangle_4)|0\rangle_5$
(the phase of $|1\rangle_4$ is changed).
For the Steane code, $\bar{X}$ is given by
$\bar{X}=X_5X_6X_7$~\cite{Gottesman}, and
the modified initial state $|\bar{1}\rangle'$ is expressed by 
$|\bar{1}\rangle'=X_2X_3X_5 |00000\rangle=|0110100\rangle$.
Hence, we have to prepare 
$a|0000000\rangle+b|0110100\rangle$ to which $\bar{M}$ is applied. 
This state is transformed from $|0\rangle_1(a|0\rangle_2+b|1\rangle_2)|00000\rangle$
by applying CNOT gates in which qubits 3 and 5 are target qubits 
while qubit 2 is the control qubit.

The nine-qubit codes can be generated in a different way, because the nine-qubit 
code is expressed by the product of three parts given by~\cite{Shor}:
\begin{eqnarray}
|\bar{0} \rangle \equiv
(|000\rangle +|111\rangle)(|000\rangle +|111\rangle)(|000\rangle +|111\rangle),
\ \ \\
|\bar{1} \rangle \equiv
(|000\rangle -|111\rangle)(|000\rangle -|111\rangle)(|000\rangle -|111\rangle).
\ \ \label{9code}
\end{eqnarray}
Each three-qubit block is a Greenberger-Horne-Zeilinger(GHZ) state. 
From 
$
|000\rangle \pm |111\rangle =\exp[ \mp i (\pi/4) X_1Y_2X_3 ] |000\rangle
$, we have
\begin{eqnarray}
|\bar{0}\rangle \!\!&\!=\!&\! 
\exp\left( -i \frac{\pi}{4} H_0^{\rm 9code}  \right)
|0...0\rangle, \ \ \\
|\bar{1}\rangle \!\!&\!=\!&\! 
\exp\left( i \frac{\pi}{4} H_0^{\rm 9code} \right)
|0...0\rangle. \ \ \ \ \ 
\end{eqnarray}
where the Hamiltonian $H_0^{\rm 9code}\equiv X_1Y_2X_3+X_4Y_5X_6+X_7Y_8X_9$ is obtained 
starting from $X_1+X_4+X_7$ by applying
operations discussed in the previous sections. 
The concrete pulse sequence is given by 
(1)$H_{XY}^{12}+H_{XY}^{45}+H_{XY}^{67}$,  
(2)$H_{XY}^{34}+H_{XY}^{56}+H_{XY}^{78}$, 
and (3) single-qubit rotations. 
Unknown state $a|0\rangle+b|1\rangle$ is encoded by 
applying $\exp\left( -i \frac{\pi}{4} H_0^{\rm 9code}  \right)$
to a changed state  $a|0...0\rangle+b|1...1\rangle$ which can be 
obtained by CNOT gate to $(a|0\rangle+b|1\rangle)|0...0\rangle$.
 
 \begin{figure}
\includegraphics[width=7cm,clip=true]{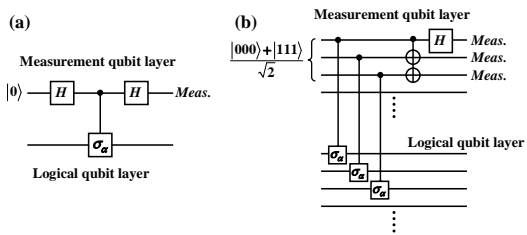}
\caption{Measurement circuit for fault-tolerant quantum computation~\cite{Nielsen}. 
In order to apply any kinds of QECC, measurement qubit is 
required for every physical qubit in the logical qubit layer.
(a) Single-qubit measurement. (b)Multi-qubit measurement. $H$ shows a Hadamard gate. } 
\label{FT}
\end{figure}

\section{qubit architecture}\label{sec:architecture}
Let us consider possible encoded qubit architectures for solid-state qubits 
controlled by local gate electrodes.
In general, solid-state qubits are fabricated on some substrate and, 
unlike optical qubits and ion trap qubits~\cite{Barreiro}, 
they cannot be moved,
being subject to the restriction that the interactions 
between qubits are limited to the nearest qubits. 
Thus, as discussed in Sec.~\ref{sec:lattice}, it is natural to set a logical qubit as a 1D array.
In order to construct various stabilizer codes, 
every qubit should be accessed by an appropriate gate electrode.
This means that a gate electrode layer should be placed along logical qubits.
Because logical qubits interact with each other in a 2D plane, the gate electrode 
layer will be constructed on or under the logical qubit layer.

\begin{figure}
\includegraphics[width=8cm]{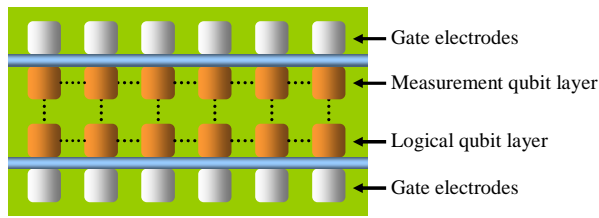}
\caption{Layered 3D QECC system. 
There are two qubit layers; a logical qubit layer and 
a measurement qubit layer.
Each qubit layer is connected 
to a gate electrode layer by which physical qubits are controlled.
Boxes show qubits and electrodes. Dot lines show qubit-qubit interaction. 
In the stabilizer coded, measurement is an indispensable process 
for encoding and decoding. 
Thus, the logical qubit layer is set close to the measurement qubit layer.} 
\label{stack}
\end{figure}

Next, let us consider a structure of measurements.
For the fault-tolerant computation, additional measurement 
circuits are required as described in Ref.~\cite{Steane2,Nielsen}.
Figure~\ref{FT} shows the measurement circuit for a single-qubit measurement 
and the multi-qubit measurement.
The multi-qubit measurement is used for stabilizer formalism (Fig.~\ref{FT}(b)).
In Fig.~\ref{FT}(b), the number of qubits in the cat state $|0...0\rangle + |1....1\rangle$, 
depends on the number of the Pauli matrices 
of the stabilizer (Fig.~\ref{FT}(b) is the case of three-qubit stabilizer).
This means that the number of ancilla qubits for the whole measurement circuit is of
the same order as that of qubits in a logical qubit layer.
Therefore, so as to avoid direct measurements and achieve the fault-tolerant computation, 
it is appropriate to set an independent qubit layer for measurements.
Because we already have a logical qubit layer, it is natural 
that the additional measurement layer should be stacked as shown in Fig~\ref{stack}.
Note that physical qubits and electrodes in Fig.~\ref{stack} are described in a abstract form.
Real qubits and electrodes are more complicated than a box.
Thus, a stacked 3D qubit system will be straightforward architecture for an effective QECC system, 
as long as we assume that the interaction between physical qubits is restricted to their neighboring qubits.
For generating the cat state of  qubits in the measurement layer,
our method shown in the previous section regarding the nine-qubit code generation  
is useful.

\begin{figure}
\includegraphics[width=8.5cm]{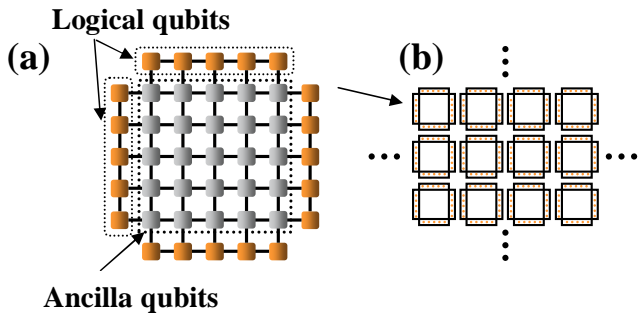}
\caption{2D qubit layout. Small box shows physical qubits. 
(a) Single logical qubit unit, which is composed of four peripheral logical qubits 
and central ancilla qubits. The four peripheral qubits are processed to be equivalent. 
They interact with logical qubits of other logical qubit units.
(b) 3$\times$4 logical qubit array where each square corresponds to the logical qubit of (a).} 
\label{2D}
\end{figure}

The stacked 3D qubit system can be applied to spin qubits and charge qubits. 
However, not all qubits can be stacked in the 3D system. 
Consider an example of standard superconducting flux qubits. 
If we stack flux qubits, the same flux penetrates stacked two qubits, resulting in confusion of 
signal between the stacked flux qubits. 
In such case, we will be able to implement a single logical qubit 
into a square form as shown in Fig.~\ref{2D}.
The 2D arrangement consists of four logical qubits placed at the peripheral
and ancilla qubits surrounded by the logical qubits.
The four logical qubits share their quantum information through 
SWAP operation in the ancilla qubits and connect 
to the four directions of the nearest logical qubits.
The ancilla qubits at the central region work for fault-tolerant measurements.

\section{Robustness against pulse errors}\label{sec:robustness} 
Since the codeword states are encoded in the twofold-degenerate ground-state
manifold $|\bar{0}\rangle$ and $|\bar{1}\rangle$ of $H_{\rm stab}$, the
robustness of this method is limited by the rate of leakage out of this
manifold.  Thus, energy non-conserving single-qubit errors---often a prevalent 
kind of errors created by a thermal bath---  are exponentially suppressed
for temperatures that are low compared to the Zeeman-splitting $\Omega$.
Hence, besides local imperfections and noise sources, unavoidable
pulse errors are likely to be the predominant cause of leakage, at
low temperatures.

In the present method, each logical qubit is constructed 
by starting from a single-qubit Hamiltonian $\sum_i\Omega_i Z_i$, and
multiplying operators like $X_1 \rightarrow X_1X_2 \rightarrow X_1\cdots X_N$.
Hence,  it is possible 
that this process makes operation errors transmit through each logical qubit. 
If we model the pulse errors by randomly distributed, unbiased, and 
uncorrelated deviations
$\delta\theta$ with $\sigma_\theta = \sqrt{\langle \delta \theta^2 \rangle}$
from the ideal angle of $\pi / 2$. The leakage from the twofold-degenerate ground-state
manifold $|\bar{0}\rangle$ and $|\bar{1}\rangle$ can then be estimated
by looking at the average of the ground state fidelity 
$\langle F (t) \rangle
\approx 1 - N_{\rm P} \sigma_\theta^2 t / (8 \mathcal{T})$, where $N_{\rm P}$
is the number of pulses in the sequence to generate $H_{\rm stab}$,
and $\mathcal{T}$ its duration~\cite{qecc-basel}.
Thus, the reduction of the number of pulses  $N_{\rm P}$ 
for generating stabilizer codes (Tables I and II) is very important. 

For the QECC scheme to succeed, the error rate of each qubit operation
should be less than 10$^{-7} \sim 10^{-5}$~\cite{Steane2,Nielsen}.
Thus the accuracy of operation pulses is crucial.
In this regard, we can also use one of many NMR techniques.
If we construct each single pulse by composite pulses, 
the accuracy of the pulse increases dramatically~\cite{Ernst}.
The composite-pulse method generalizes the concept of spin echo, and
has already been applied in the field of quantum computation
to greatly improve both single-qubit rotations and CNOT operations~\cite{Haffner,Molmer,Hill,Torosov}.
As the number of pulses $N_{\rm P}$ decreases and the dephasing time $T_2$ increases, 
more accurate composite pulses can be implemented, resulting in the success of QECC scheme.

\section{Summary and Conclusions}\label{sec:conclusion}
In summary, we showed how to produce stabilizer Hamiltonians starting from 
natural two-body Hamiltonians by using appropriate pulse 
sequences. 
We demonstrated our method by using typical codes: the nine-qubit code, the five-qubit code and 
the Steane code. 
The key method of finding the pulse sequence is to inversely trace 
the derivation process from the stabilizer Hamiltonian to the single-qubit Hamiltonian.
We also showed how to generate encoded states without using measurements.
Stabilizer Hamiltonians are important for preserving encoded states
as ground states of the system.
Effective preparation of stabilizers is considered to be critical to the succeed of QECC.

Many important experiments have been performed to enlarge 
coherence time in solid-state qubits~\cite{Steffen}.
The criteria for the realization of quantum computing 
is whether a sufficient number of quantum operations can be carried out
during a given coherence time.
Thus, manipulation speed of each quantum operation 
is one of the most important factors for practical quantum computing.
Considering the fact that a quantum computer exceeds a digital computer 
only in several fields such as search algorithm, 
it will be natural to embed a quantum computer as a part of a digital computer system.
Moreover, as in the present experiments, a quantum circuit will be operated by 
a digital computer.
Although the speed of a single processing unit of 
a commercial digital computer seems to become saturated, 
performance of digital computers 
will continue to increase by parallel processing. 
Accordingly, it is expected that the manipulation speed of a pulse 
sequence will also increase. 
Therefore, the approach presented in this paper enables 
faster quantum operations by using the cutting-edge 
technology of computer science.
How to achieve an appropriate and smooth connection between a quantum computer and a 
digital computer will be a future problem. 

\acknowledgements
We would like to thank C. Bruder, V. M. Stojanovi\'c, D. Becker,
A. Nishiyama, K. Muraoka, S. Fujita, H. Goto, Y.X. Liu and F. Nori  
for discussions.

\appendix 
\section{Extraction part of other type of Hamiltonian}\label{appendixA}
Here, we show how to extract $H_0$ or interaction parts  
from a Hamiltonian that includes three Pauli matrices $X$, $Y$ and $Z$.
This situation appears for Eq.(\ref{app2}) with $\omega_i^{\rm rf}\neq 2\Omega_{0i}$  or Ref.~\cite{Liu}
in which the Hamiltonian is given by
\begin{equation}
H=\sum_i \left[  \omega_{i} Z_i +\epsilon_i X_i  \right]
\nonumber \\
+\sum_{i<j}\frac{J_{ij}}{2}
[X_iX_j +Y_iY_j].
\label{app3}
\end{equation}
For this Hamiltonian, one more step is required to obtain 
both $H_0$ and an interaction part. 
When the method of Sec.~\ref{sec:extractH_0} is applied,
we obtain $H_{\rm eff}=\sum_{k,i} [  4 \omega_{i} Z_i^{(k)} +2 \epsilon_i X_i^{(k)} ]$ 
after using Eq.(\ref{ABAB}).
When the method of Sect.~\ref{sec:extractH_int} is applied,
we obtain $H_{\rm eff}'=\sum_{k}[\sum_{i} 4Z_i^{(k)}+ \sum_i H_{XY}^{23(k)}]$.
In both cases, extra $\sum_{i} Z_i$ term remains. 
Therefore, we need one more step to delete $\sum_{i} Z_i$ term such as 
$e^{ -i\tau H_{\rm eff}}e^{-i\pi \sum_i X_i/2} e^{ -i\tau H_{\rm eff}} e^{i\pi \sum_i X_i/2}$, 
and $e^{ -i\tau H_{\rm eff}'}e^{-i\pi \sum_i X_i/2} e^{ -i\tau H_{\rm eff}'} e^{i\pi \sum_i X_i/2}$.

\section{Perturbation terms in BCH formula}\label{appendixB}
Here, we show the first-order perturbation terms that appear during the process of extracting $H_{XY}^{23}$ and 
a single-qubit Hamiltonian $H_0$ from the original Hamiltonian $H_q$ 
discussed in Sec.~\ref{sec:extract}.

The first-order perturbation terms from the ideal Hamiltonian $H_{XY}^{23}$ in five qubits are 
given by
\begin{equation}
H_{\rm pert}\approx 
-\sum_k 2 \tau \{ P^{(k)}+Q^{(k)}+R^{(k)} \}
\end{equation}
where
\begin{eqnarray}
P^{(k)} &=& J_{12}^{(k-1)}\Omega_1^{(k-1)}Y_2^{(k-1)} Z_1^{(k-1)} \nonumber \\
&+& J_{34}^{(k-1)}\Omega_4^{(k-1)}Y_3^{(k-1)} Z_4^{(k-1)} \nonumber \\
&+&  J_{45}^{(k-1)}\Omega_4^{(k-1)}Y_5^{(k-1)} Z_4^{(k-1)}
\nonumber \\
&+& S_{\beta,1}^{(k)}+S_{\beta,4}^{(k)}
+ S_{\alpha,2}^{(k)}+S_{\alpha,3}^{(k)}+S_{\alpha,5}^{(k)},\\
Q^{(k)} &=& J_{23}^{(k)} 
   [\Omega_2^{(k)}Y_3^{(k)} Z_2^{(k)} 
  + \Omega_3^{(k)}Y_2^{(k)} Z_3^{(k)}],
\\
R^{(k)} 
&=& J_{12}^{(k)}\Omega_2^{(k)}Y_1^{(k)} Z_2^{(k)} 
   +J_{34}^{(k)}\Omega_3^{(k)}Y_4^{(k)} Z_3^{(k)} 
\nonumber \\
&+& J_{45}^{(k)}\Omega_5^{(k)}Y_4^{(k)} Z_5^{(k)}
\nonumber \\
&+& S_{\alpha, 1}^{(k+1)} +S_{\alpha, 4}^{(k+1)} 
  + S_{\beta, 2}^{(k+1)} +S_{\beta, 3}^{(k+1)}+S_{\beta, 5}^{(k+1)} 
  \nonumber \\
\end{eqnarray}
where 
\begin{eqnarray}
S_{\alpha,i}^{(k)}&=& J_{ii}^{(k,k-1)}\Omega_{i}^{(k)} Y_{i}^{(k-1)}Z_{i}^{(k)}
\\
S_{\beta,i}^{(k)} &=& J_{ii}^{(k,k-1)}\Omega_{i}^{(k-1)} Y_{i}^{(k)}Z_{i}^{(k-1)},
\end{eqnarray}
Thus, the perturbation terms can be described by
\begin{equation}
|| H_{\rm pert} || \approx 10 \tau N_{\rm qubit} J\Omega,
\label{appendix1}
\end{equation}
where $N_{\rm qubit}$ is the total number of qubits in a circuit. 
$N_{\rm qubit}$ is expressed by $N_{\rm qubit}=N_{\rm logic}N_{\rm phys}$
with the number of logical qubits $N_{\rm logic}$ and that of 
physical qubits in a logical qubit $N_{\rm phys}$. 

The first-order perturbation term to obtain the 
single-qubit Hamiltonian is given by
\begin{eqnarray}
F_a^{(k)}\!\!&=& \!\!  J_{12}^{(k)}\Omega_1^{(k)} Y_2^{(k)} Z_1^{(k)}
+[J_{23}^{(k)}Y_2^{(k)}+ J_{34}^{(k)}Y_4^{(k)}]\Omega_3^{(k)}  Z_3^{(k)}
\nonumber \\
&+&
J_{45}^{(k)}Y_4^{(k)}\Omega_5^{(k)}  Z_5^{(k)},
\nonumber \\
F_b^{(k)}\!\!&=&\!\!
[J_{12}^{(k)}Y_1^{(k)}+ J_{23}^{(k)}Y_3^{(k)}]\Omega_2^{(k)}  Z_2^{(k)}
\nonumber \\
&+&
[J_{34}^{(k)}Y_3^{(k)}+ J_{45}^{(k)}Y_5^{(k)}]\Omega_4^{(k)}  Z_4^{(k)},
\nonumber \\
W_a^{(k)}\!\!&=&\!\!
S_{\alpha,1}^{(k)}+S_{\alpha,3}^{(k)}+S_{\alpha,5}^{(k)}
+S_{\beta,1}^{(k+1)}+S_{\beta,3}^{(k+1)}+S_{\beta,5}^{(k+1)}
\nonumber \\
W_b^{(k)}\!\!&=&\!\!
S_{\alpha,2}^{(k)}+S_{\alpha,4}^{(k)} 
+S_{\beta,2}^{(k+1)}+S_{\beta,4}^{(k+1)}
\nonumber
\end{eqnarray}
\begin{eqnarray}
H_{\rm pert}&=&2\tau \{ \sum_k (-)^{k+1} (F_a^{(k)}+F_b^{(k)}) \}
\nonumber \\
&+& \tau [8+(-)^k 2] (-W_a^{(k)}+W_b^{(k)})
\end{eqnarray}
Thus the perturbation terms can be estimated by
\begin{equation}
|| H_{\rm pert} || \approx 20 \tau N_{\rm qubit} J\Omega.
\label{appendix2}
\end{equation}

Eqs.~(\ref{appendix1}) and (\ref{appendix2}) show that the number of connected qubits should be small 
so that the perturbation terms do not affect the main terms, 
even when we reduce the perturbation terms by using Eq.~(\ref{ABAB}).
Therefore, instead of connecting all qubits by 
always-on Hamiltonian, it is better to divide qubits into several blocks 
such that the blocks are connected by some kinds of switching mechanism~\cite{Zagoskin,Grajcar,Niskanen,Ashhab}.


\end{document}